\documentclass[5p]{elsarticle}

\usepackage{hyperref}
\usepackage[T1]{fontenc}
\usepackage[utf8]{inputenc}
\usepackage{textcomp} 
\usepackage{lmodern} 
\usepackage{multirow}
\usepackage{amsmath}
\usepackage{amsfonts}
\usepackage{breakurl} 
\usepackage[switch]{lineno} 

\usepackage[colorinlistoftodos]{todonotes}

\definecolor{hlcolor}{RGB}{225, 25, 0}

\graphicspath{{./figures/}}
\biboptions{sort&compress}

\hypersetup{
    breaklinks = true,
    colorlinks = true,
    pdftitle = {}, 
    pdfauthor = {},
    pdfkeywords = {},
    linkcolor = blue,
    citecolor = blue,
    filecolor = black,
    urlcolor = magenta
}

\makeatletter 
\def\@author#1{\g@addto@macro\elsauthors{\normalsize%
    \def\baselinestretch{1}%
    \upshape\authorsep#1\unskip\textsuperscript{%
      \ifx\@fnmark\@empty\else\unskip\sep\@fnmark\let\sep=,\fi
      \ifx\@corref\@empty\else\unskip\sep\@corref\let\sep=,\fi
      }%
    \def\authorsep{\unskip,\space}%
    \global\let\@fnmark\@empty
    \global\let\@corref\@empty  
    \global\let\sep\@empty}%
    \@eadauthor={#1}
}
\makeatother

\begin{document}
\begin{sloppypar} 

\title{First-principles study of the magnetic anisotropy of ultrathin B-, C-, and N-doped FeCo films}

\author{Joanna Marciniak}
\author{Miros\l{}aw Werwi\'nski}
\author{Justyna Rych\l{}y-Gruszecka\corref{cor1}}
\ead{justyna.rychly-gruszecka@ifmpan.poznan.pl}
\cortext[cor1]{Corresponding author}
\address{Institute of Molecular Physics, Polish Academy of Sciences,  M. Smoluchowskiego 17, 60-179 Pozna\'n, Poland}

\begin{abstract}
Iron-based layered systems are of great interest because of their ability to tune effective material parameters such as magnetic anisotropy energy (MAE).
The influence of the crystallographic structure of Fe, its thickness, and the presence of other layers above and below the Fe layer on magnetic parameters, such as the MAE of the studied system, is an intriguing and important topic from an application point of view. 
Here, we present a density functional theory (DFT) study of the magnetic anisotropy of nine-monolayer Fe, FeCo, and FeCo films with B, C, and N dopants placed in octahedral interstitial positions. 
The theoretical study is based on calculations using the full-potential local-orbital code FPLO and the generalized gradient approximation. 
The chemical disorder in the FeCo layers was modeled using the virtual crystal approximation. 
The structures of the layers were subjected to optimization of the geometry of the interlayer spacings and the neighborhood of the dopant sites. 
We determined the local magnetic moments and the excess charge at each layer position. 
We also identified the influence of dopant atoms on the magnetic properties of FeCo layers, such as magnetization and magnetic anisotropy.
\end{abstract}


\maketitle

\section{Introduction}
Permanent magnets are essential in many consumer and industrial products that convert energy. 
Rare earth magnets, particularly those based on neodymium and samarium, typically produce the highest energy product. 
Of the various permanent magnets reported to date, \mbox{NdFeB} has the highest magnetic efficiency. 
Recently, there has been growing concern about the volatility of the rare earth market, which manifested itself in the so-called rare earth crisis of 2011 \cite{coey2012permanent}. 
Over the past decade, the rapidly rising and volatile prices of rare earth metals have prompted intensive research efforts worldwide to develop alternative materials for permanent magnets, especially those with very low or zero rare earth content.
FeCo has the highest saturation magnetization of all transition metals and their alloys. 
Still, due to its cubic crystal structure (bcc), it has an extremely low magnetic anisotropy, making it a typical soft magnetic material, which has long been unsuitable for permanent magnets. 
Burkert et al. \cite{burkert2004giant} performed first-principles calculations using a virtual atom model to theoretically predict that FeCo with a tetragonal crystal structure (bct) would exhibit a giant magnetic anisotropy constant of up to $10^{7}~\mathrm{J\,m}^{-3}$. 
In a FeCo alloy with lattice parameter ratio $c/a \approx 1.2$ indication of high tetragonal deformation, the magnetic anisotropy energy (MAE) was predicted to reach $0.7-0.8$ meV\,atom$^{-1}$.
However, the experimental growth of Fe-Co alloys with such high tetragonal distortion is challenging because their natural structure is bcc (up to about 70\% Co concentration).
Subsequent theoretical studies have corrected the theoretical MAE to values very close to the experimental ones, using a more realistic treatment of the chemical disorder present in Fe-Co alloys \cite{neise2011effect, turek2012magnetic}.
To solve this problem, tiny interstitial atoms (B, C, N, etc.) placed on the $c$-axis of FeCo can stretch the unit cell and stabilize the bct structure.
As for FeCo, there are studies on adding interstitial elements 
B~\cite{reichel2015soft, coene1991magnetocrystalline, chen2000soft}, 
C~\cite{reichel2015lattice, DelczegStabilizationFeCoCPhysRevB2014, marciniak2023structural}, 
N~\cite{hasegawa2019stabilisation, hasegawa2021challenges, zhao2016large}.

In the race to miniaturize electronic devices, modern science has reached the ultimate level of monoatomic layers.
The limit for magnetic tunnel junctions with perpendicular magnetic anisotropy is about twenty atomic monolayers \cite{watanabe2018shape, snarski2022magnetic}.
Layered systems are particularly interesting for their ability to tune effective material parameters such as MAE. 
Among such systems, iron-based layered systems are of considerable interest \cite{perumal2008particulate, hammar2022theoretical, miura2013origin, ito2020epitaxial}.
An intriguing and important topic from the point of view of applications is the influence of the crystallographic structure of Fe, its thickness, and the presence of other layers above and below the Fe layer on magnetic parameters such as the MAE of the studied system.
\begin{figure}[t]
    \centering
    \includegraphics[clip,width=0.9\columnwidth]{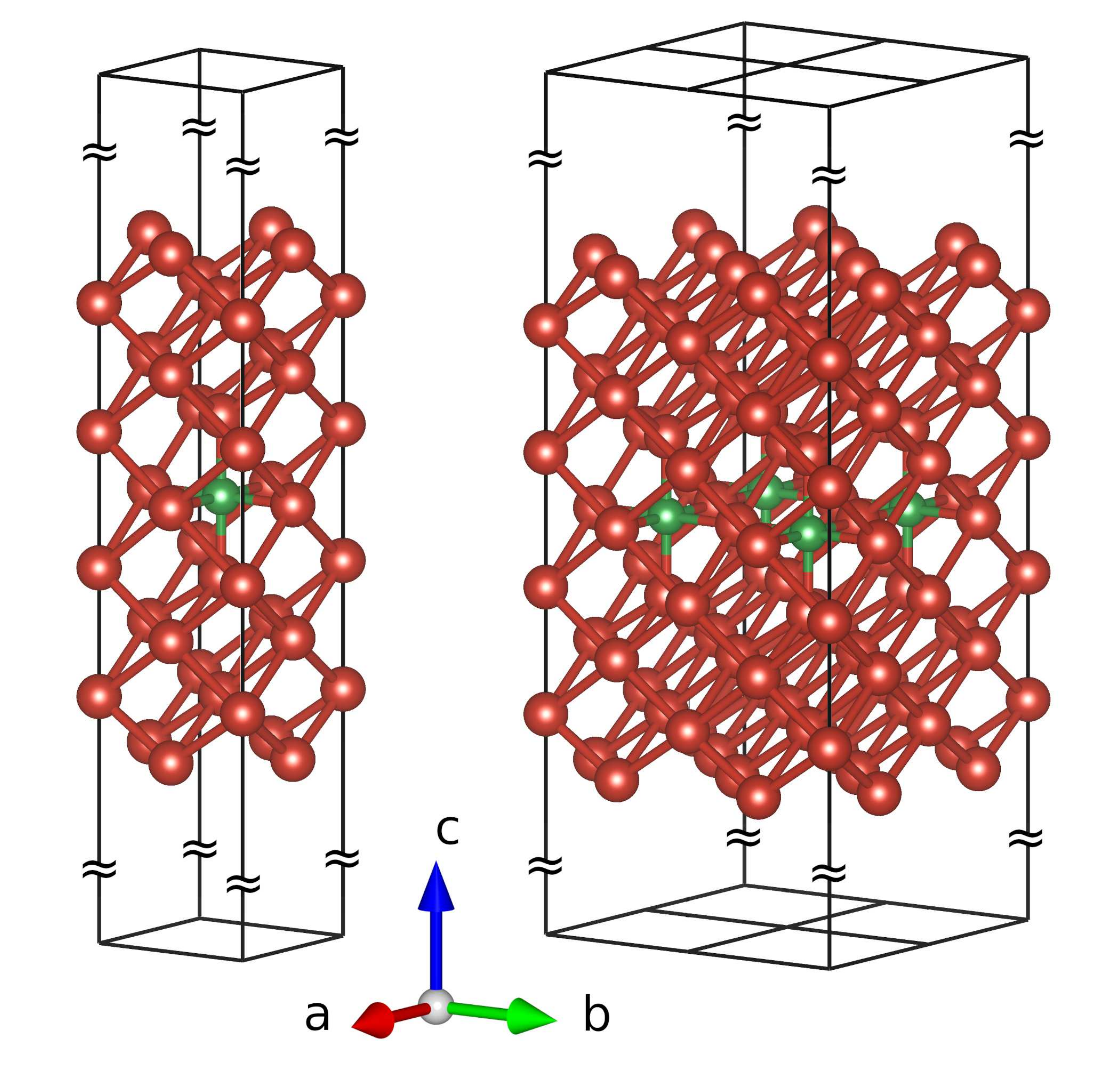}   
    \caption{
    A unit cell of a 9-atomic monolayer thick Fe$_{0.7}$Co$_{0.3}$ thin film, with the dopant (B or C or N) located in an octahedral interstitial position.
    A single computational cell is shown on the left, and a multiplied $2\times2$ cell on the right.
    The unit cell is obtained after geometry optimization. 
    The red spheres represent the Fe$_{0.7}$Co$_{0.3}$ virtual atoms, while the green sphere represents the dopant atom. 
    A surrounding vacuum (52~\AA{}) was applied in the computational cells to obtain non-interacting thin films.
    To improve the visibility of the film's atoms, the height of the vacuum used has been significantly reduced in the figures.
}
\label{fig:structure}
\end{figure}

Here, we present a theoretical study of the magnetic anisotropy of FeCo thin films with B, C, and N dopants located in octahedral interstitial positions. 
The theoretical study is based on calculations using the full-potential local-orbital electronic structure code (FPLO) and the generalized gradient approximation. 
The chemical disorder in FeCo layers is modeled using the virtual crystal approximation (VCA). 
The layer structures are subjected to geometry optimization of the interlayer distances and the vicinity region of the dopant sites. 
We determine the local magnetic moments and excess charge at each atomic position. 
We identified the dopant atoms' effect on the FeCo films' magnetic properties, such as magnetization and magnetic anisotropy.

\section{Calculations' details}
\begin{figure}[t]
    \centering
    \includegraphics[clip,width=6.0cm]{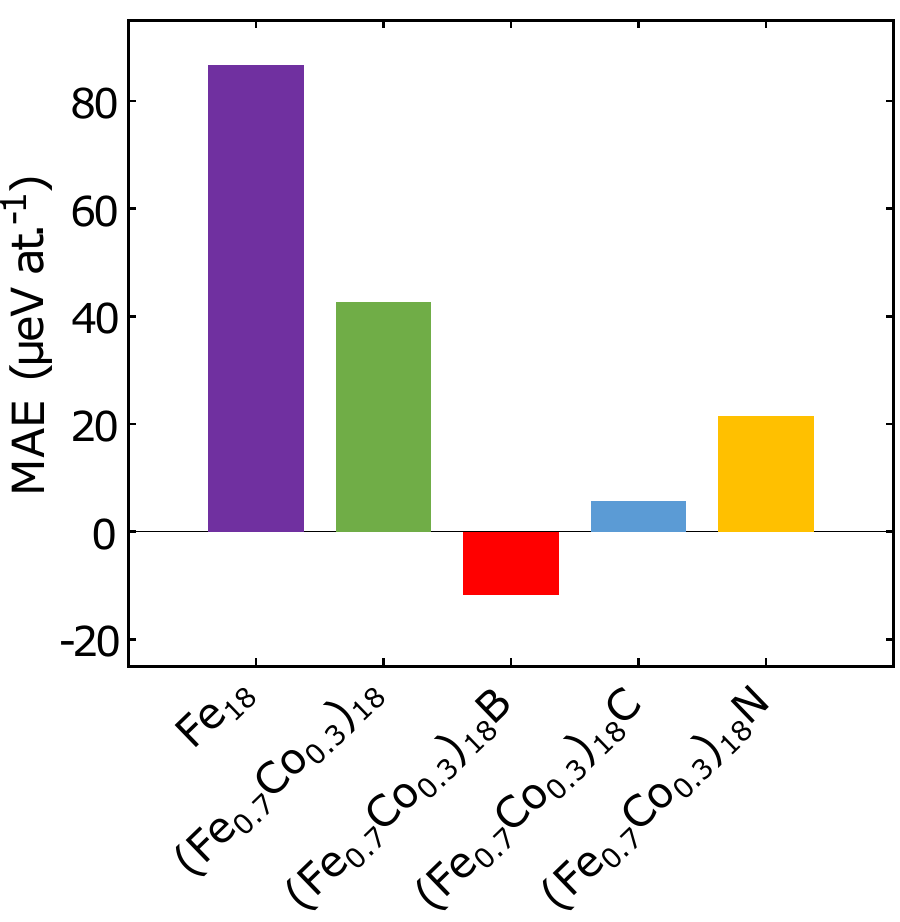}
    \caption {Magnetic anisotropy energy (MAE) calculated for 9 atomic monolayer thick films. From left to right: 
    pure Fe (shown as a purple bar); 
    pure Fe$_{0.7}$Co$_{0.3}$ alloy (shown as a green bar); 
    Fe$_{0.7}$Co$_{0.3}$ with a B dopant atom (shown as a red bar); 
    Fe$_{0.7}$Co$_{0.3}$ with a C dopant atom (shown as a blue bar); 
    Fe$_{0.7}$Co$_{0.3}$ with an N dopant atom (shown as a yellow bar). 
    The dopant atoms are in the octahedral interstitial position. 
    The DFT calculations were performed with the full-potential local-orbital (FPLO-18) code \cite{koepernikFullpotentialNonorthogonalLocalorbital1999}. 
    The exchange-correlation functional in the Perdew-Burke-Ernzerhof (PBE) parametrization was used \cite{perdewGeneralizedGradientApproximation1996}.
    }
    \label{fig:MAE} 
\end{figure}

In this study, we examined the structural and magnetic properties of ultrathin (9 atomic monolayers thick) films of Fe$_{0.7}$Co$_{0.3}$ alloys with a dopant (of B, C, and N atoms) located in octahedral interstitial position, see Fig.~\ref{fig:structure}, using density functional theory (DFT) calculations.
To compare the effect of the dopant, placed in the octahedral gap, on the obtained magnetic properties of the structures, we also performed calculations of thin films of pure iron-cobalt alloy with a given concentration (70\%~Fe and 30\%~Co) and a thin film of iron without doping. 
Altogether, calculations of five different structures were performed.

The structures were prepared in cells with symmetry $P4/mmm$, where the lattice constant equals $a=4.0~\AA{}$. 
In the computational cells, to remove the bulk periodicity in the indicated direction and to model the thin film, a vacuum of 52~\AA{} was added, giving a height of the computational cell equal to 64\,\AA{}. 
In the presented calculations, the VCA approach was used, as implemented in FPLO 18.00-52 code, to simulate a disordered 30\%~Co alloy.
Wyckoff's positions were optimized in each considered system using a scalar-relativistic approach with spin polarization. 
Atomic positions in [001] were optimized by forces optimization on atoms, with convergence criterion set as 10$^{-3}$~eV\,\AA{}$^{-1}$.

A mesh of $k$-points was set as 65~$\times$~65~$\times$~5, and the convergence criteria were used 10$^{-8}$~Ha and 10$^{-7}$ for energy and charge density, respectively.
The integration over the Brillouin zone was performed using the tetrahedron method. 
The DFT calculations were performed using the full-potential local-orbital approach implemented in the FPLO~18.00-52 code \cite{koepernikFullpotentialNonorthogonalLocalorbital1999}. 
Using the Perdew-Burke-Ernzerhoff (PBE) exchange-correlation potential \cite{perdewGeneralizedGradientApproximation1996}, scalar-relativistic calculations were performed, followed by 1 iteration of full-relativistic calculations.
MAE was determined by the formula: 
\begin{equation}\label{eq:MAE}
    \mathrm{MAE} = E(\theta=0^\circ) - E(\theta=90^\circ),
\end{equation}
where $\theta$ is the angle between the magnetization direction and the 
$c$ axis.
Crystal structures were visualized using the VESTA program \cite{mommaVESTAThreedimensionalVisualization2008}.

\section{Results and Discussion}

\begin{table*}[t]
\center
\caption{
Total thicknesses of the 9-atomic monolayer thick Fe$_{0.7}$Co$_{0.3}$ and Fe$_{0.7}$Co$_{0.3}$-X thin films after interlayer spacing optimization, together with distances between the central atomic monolayer and the first and second-nearest monolayer.
The double values for the atoms of the first nearest monolayer are due to the presence of dopant atoms in the central layer (see Fig.~\ref{fig:structure}).
The $a$ lattice parameter of the computational unit cell was set at 4.00~\AA{} ($\sqrt{2}\times 2.83$~\AA{}).\\
}
\label{tab-thickness-change}
\begin{tabular}{c c c c c}
        \hline
        \hline
              &  (Fe$_{0.7}$Co$_{0.3}$)$_{18}$    & (Fe$_{0.7}$Co$_{0.3}$)$_{18}$B    & (Fe$_{0.7}$Co$_{0.3}$)$_{18}$C    &  (Fe$_{0.7}$Co$_{0.3}$)$_{18}$N\\
        \hline
        total thickness (\AA{}) &  11.27    &  11.81    & 11.74    &  11.75\\
        1$^{st}$ layer (\AA{})   &  1.42    &  1.58/1.86    & 1.59/1.80  &  1.60/1.80\\
        2$^{nd}$ layer (\AA{})  & 2.83    &  3.09   &  3.06   &  3.07\\
        \hline
        \hline
    \end{tabular}
\end{table*}

\begin{figure*}[t]
    \centering
    \includegraphics[clip,width=0.9\textwidth]{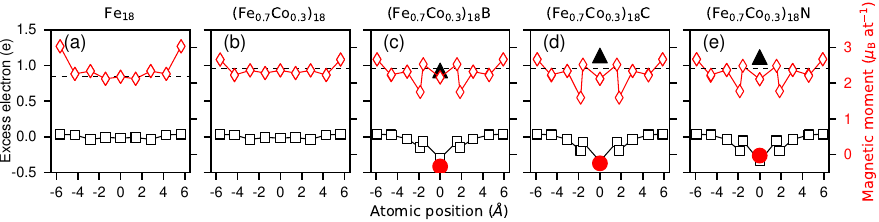}
    \caption {Excess charge (charge transfer) at each atomic position in the calculated thin films presented as black empty squares and big black full triangles for dopant atoms. Local spin magnetic moments at each atomic position in the calculated thin films presented as red empty diamonds and big red full circles for dopant atoms. 
    Considered 9-atomic monolayer thick films of 
    (a) pure Fe; 
    (b) pure Fe$_{0.7}$Co$_{0.3}$ alloy; 
    (c) Fe$_{0.7}$Co$_{0.3}$ with a B dopant atom; 
    (d) Fe$_{0.7}$Co$_{0.3}$ with a C dopant atom; 
    (e) Fe$_{0.7}$Co$_{0.3}$ with an N dopant atom. 
    The dopant atoms in (c-e) are in the octahedral interstitial position. 
    Calculations were performed with the PBE exchange-correlation potential \cite{perdewGeneralizedGradientApproximation1996} in scalar-relativistic formalism using the FPLO18 code \cite{koepernikFullpotentialNonorthogonalLocalorbital1999}.
    }
    \label{fig:e_m_at} 
\end{figure*}

\begin{figure}[t]
    \centering
    \includegraphics[clip,width=1.0\columnwidth]{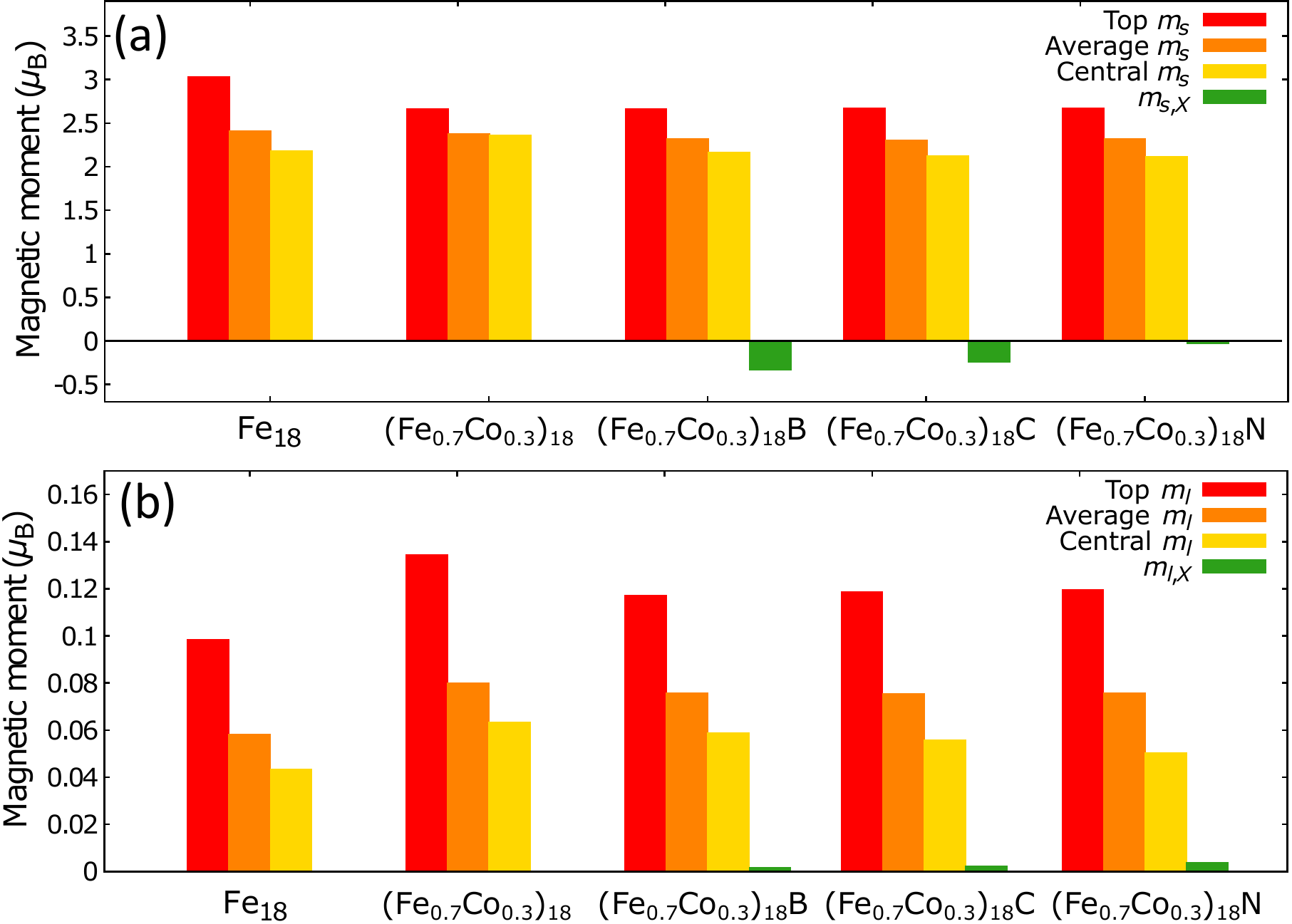}
    \caption {
    The calculated (a) spin and (b) orbital magnetic moment for all considered 9-atom layers
    (from left to right: pure Fe; 
    pure Fe$_{0.7}$Co$_{0.3}$ alloy; 
    Fe$_{0.7}$Co$_{0.3}$ with B-, C-, or N-atom, located in the octahedral interstitial position, respectively. 
    The red, yellow, and orange bars represent the (a) spin (b) orbital magnetic moment calculated for the outer, central, average atom (Fe in case of pure Fe thin film or Fe$_{0.7}$Co$_{0.3}$ for the rest of the structures), respectively; the green bar represents a (a) spin (b) orbital magnetic moment for the dopant atom. 
    The DFT calculations were performed with the full-potential local-orbital (FPLO-18) code \cite{koepernikFullpotentialNonorthogonalLocalorbital1999}. The exchange-correlation functional in the Perdew-Burke-Ernzerhof (PBE) parametrization was used \cite{perdewGeneralizedGradientApproximation1996}). 
    }
    \label{fig:ms_ml} 
\end{figure}

\begin{figure}[t]
    \centering
    \includegraphics[clip,width=0.9\columnwidth]{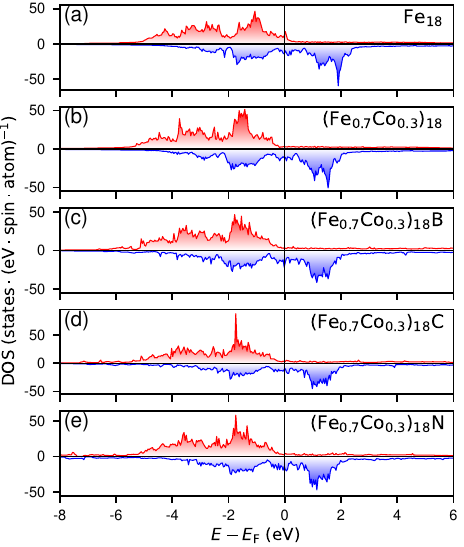}
    \caption {
    The calculated total density of states (DOS) for the considered 9-atom thick layers of:
    (a) pure Fe; 
    (b) pure Fe$_{0.7}$Co$_{0.3}$ alloy; 
    (c) Fe$_{0.7}$Co$_{0.3}$ with a B dopant atom; 
    (d) Fe$_{0.7}$Co$_{0.3}$ with a C dopant atom; 
    (e) Fe$_{0.7}$Co$_{0.3}$ with an N dopant atom. 
    The dopant atoms in (c-e) are in the octahedral interstitial position. 
    The DFT calculations were performed with the full-potential local-orbital (FPLO-18) code \cite{koepernikFullpotentialNonorthogonalLocalorbital1999}. The exchange-correlation functional in the Perdew-Burke-Ernzerhof (PBE) parametrization was used \cite{perdewGeneralizedGradientApproximation1996}. 
    }
    \label{fig:DOS_1} 
\end{figure}

\begin{table*}[t]
\caption{Averaged spin ($m_s$) and orbital ($m_l$) magnetic moments, together with magnetocrystalline anisotropy energies (MAE) of 9-monolayer thick films of Fe, Fe$_{0.7}$Co$_{0.3}$, and Fe$_{0.7}$Co$_{0.3}$X models, where X~=~B, C, N. 
The obtained results were compared with literature values for similar systems.
}
\label{tab-comparison-MAE-moments}
\center
\begin{tabular}{c c c c c c c c}
        \hline
        \hline
              System  & Form & Method & Reference &    $m_s$  & $m_l$  & MAE  & MAE  \\
              &  &  &    &   ($\mu_\mathrm{B}$\,at$^{-1}$) & ($\mu_\mathrm{B}$\,at$^{-1}$) & ($\mu$eV\,at$^{-1}$) & (MJ\,m$^{-3}$) \\              
        \hline
       Fe &  9-monolayers &  & this work  &  2.41 & 0.058 & 87 & 1.38 \\
       Fe &  bulk         &  & \cite{snarski2022magnetic}  & 2.16  & 0.043 & 0 & 0 \\       
       Fe$_{0.7}$Co$_{0.3}$  &  9-monolayers & VCA  & this work  &  2.38 & 0.080 & 43 & 0.68 \\
       Fe$_{0.7}$Co$_{0.3}$  & bulk & VCA/CPA  &  \cite{DelczegStabilizationFeCoCPhysRevB2014}  &  2.42/2.43 & -- & 54/27 & -- \\      
       (Fe$_{0.7}$Co$_{0.3}$)$_{18}$B  &  9-monolayers & VCA  & this work  &  2.18 & 0.076 & -13 & -0.18 \\
        (Fe$_{0.65}$Co$_{0.35}$)$_{24}$B & bulk & CPA &  \cite{reichel2015soft}  & 2.16 & 0.064 & 23 & 0.31 \\
        Fe$_{67}$Co$_{18}$B$_{15}$  & >100~nm film & expt. &  \cite{chen2000soft}  & $\sim1.7$ & -- & $\sim0$ & $\sim0$ \\
       (Fe$_{0.38}$Co$_{0.62}$)$_{0.98}$B$_{0.02}$ & 20~nm films & expt.  & \cite{reichel2015soft}   & 1.9 $\pm$ 0.2 & -- & -- & 0.4 $\pm$ 0.2\\         
       (Fe$_{0.7}$Co$_{0.3}$)$_{18}$C  &  9-monolayers & VCA  & this work  & 2.16 & 0.075 & 6 & 0.09 \\
       (Fe$_{0.7}$Co$_{0.3}$)$_{8}$C   & bulk & VCA/CPA & \cite{DelczegStabilizationFeCoCPhysRevB2014}  &   1.88/1.85 & -- & 77/81 & 1.10/-- \\
       (Fe$_{0.69}$Co$_{0.31}$)$_{16}$C  & bulk & supercells &  \cite{marciniak2023structural}  & 2.35 & -- & 65 & 0.95 \\
       (Fe$_{0.4}$Co$_{0.6}$)$_{0.98}$C$_{0.02}$ & 5~nm films & expt.  & \cite{reichel2015lattice}   & 2.0 & -- & -- & 0.80 $\pm$ 0.15\\          
       (Fe$_{0.7}$Co$_{0.3}$)$_{18}$N  &  9-monolayers & VCA  & this work  & 2.19 & 0.076 & 22 & 0.33 \\
       (Fe$_{0.75}$Co$_{0.25}$)$_{16}$N$_2$  & bulk & supercells &  \cite{zhao2016large}  & 1.96 & -- & -- & 1.5 \\
       (Fe$_{0.5}$Co$_{0.5}$)$_{89}$V$_9$N$_2$ & 20~nm film & expt. & \cite{hasegawa2019stabilisation}  & --  & --  & -- & 1.24 \\ 
        \hline
        \hline
    \end{tabular}
\end{table*}

We started the calculations for the considered structures by optimizing the $z$ positions of all atomic sites with a fixed value of the lattice parameter $a$.
The total thicknesses of the Fe-Co thin films and the distances of the nearest and second-nearest atomic monolayers to the central layer were collected in Table~\ref{tab-thickness-change}.
As we can see, the film's total thickness increases due to doping. The lattice stretching in the $z$ direction is caused by the dopant atom located in the octahedral hole.

After that, we calculated the magnetic anisotropy energy of FeCo thin films (9 atoms thick) with B, C, and N dopants in octahedral interstitial position in the center of the layer. For comparison, we have calculated a pure thin film of an iron-cobalt alloy of a given concentration and a thin film of iron. The MAE for the above structures is shown in Fig.~\ref{fig:MAE}.
For the structures of a given thickness, we can see that pure Fe layers have the strongest tendency to be magnetized perpendicular to the surface; in the case of the Fe$_{0.7}$Co$_{0.3}$ alloy, the MAE decreases almost by half, while for the thin films doped with N, C, and B elements the MAE becomes smaller and smaller. In the case of the Fe$_{0.7}$Co$_{0.3}$ with a B dopant atom, the MAE changes sign, which shows the change in the tendency of this structure and its preference to be magnetized in-plane (and not out-of-plane as was the case for the other systems). 
%

While in-plane anisotropy resulting from B doping appears uncommon in comparison to other cases, it is frequently obtained in FeCoB thin films through experimental means~\cite{kipgen2012plane}.
Based on earlier experimental work, the orientation of magnetization in FeCoB films can be altered through various technological methods, including post-deposition annealing in a magnetic field or application of mechanical stress~\cite{kipgen2012plane}.
The direction of magnetization is influenced by various factors, such as the concentration of individual elements, the type and thickness of the substrate and overlay, and the thickness of the magnetic film.
The calculated result for the ultrathin Fe$_{0.7}$Co$_{0.3}$-B film implies that it has an intrinsic property of in-plane magnetic anisotropy, unaffected by the aforementioned factors.

We also have conducted a detailed study of the charge and the spin magnetic moment for all the systems under study and present the results in Fig.~\ref{fig:e_m_at}. We first plotted (for comparison) the charge transfer at each atom of pure Fe and pure Fe$_{0.7}$Co$_{0.3}$ alloy through the thickness of their layers using black empty squares connected by a black line (see Fig.~\ref{fig:e_m_at}(a) and (b), respectively). Only negligibly small charge oscillations can be observed for these single-component layers, appearing due to the presence of surfaces. By introducing a dopant into the octahedral gap located in the center of a thin FeCo alloy layer, we can observe the transfer of a negative charge by the dopant atom at the expense of the alloy atoms in its vicinity (see Fig.~\ref{fig:e_m_at}(c-e) for dopants with B, C, and N, respectively, where the additional charge on the dopant has been visualized with a full black triangle).

We then plotted, using red empty diamonds connected by a red line, the local spin magnetic moments at each atomic position in the calculated thin films (see Fig.~\ref{fig:e_m_at}). 
It can be observed that the introduction of surfaces into the structure of pure magnetic layers without dopants increases the spin magnetic moment near their surface. 
This is particularly evident for a thin layer of pure Fe (see Fig.~\ref{fig:e_m_at}(a)), a slightly less pronounced increase in the magnetic moment is obtained for a thin layer of Fe$_{0.7}$Co$_{0.3}$ alloy (see Fig.~\ref{fig:e_m_at}(b)). 
For comparison, the thin dashed line has been used to indicate the values of the spin magnetic moment for solid Fe ($m_{\rm{s}}~=~2.2~\mu_{\rm{B}}\rm~{at}^{-1}$, marked in Fig.~\ref{fig:e_m_at}(a)) and for solid Fe$_{0.7}$Co$_{0.3}$ ($m_{\rm{s}}~=~2.42~\mu_{\rm{B}}\rm~{at}^{-1}$, marked in Fig.~\ref{fig:e_m_at}(b-e)).
By introducing the dopant into the octahedral gap in the center of the Fe$_{0.7}$Co$_{0.3}$ alloy thin film, we can observe a varying spin magnetic moment on the alloy atoms near the dopant. The dopant atom acquires a small negative spin magnetic moment (see Fig.~\ref{fig:e_m_at}(c-e) for dopants with B, C, and N, respectively, where the spin magnetic moment on the dopant has been visualized with a large red solid circle). 
The values of spin and orbital magnetic moments on the dopants obtained in the calculations are summarized in Table~\ref{tab-results}. 
Averaged values of magnetic moments at Fe/Co position and magnetic moments from literature are also summarized in Table~\ref{tab-comparison-MAE-moments}. As can be seen, our results are comparable to those obtained in experiments on thin films and calculations for solids.

To make the spin magnetic moment results easier to read, we have presented the obtained results in the form of a bar chart, where the averaged spin magnetic moment is shown with an orange bar, the spin magnetic moment for the central (outer) atom is shown with a yellow (red) bar, and the spin magnetic moment for the dopant atom is shown with a green bar (see Fig.~\ref{fig:ms_ml}(a)). Similarly, we presented the results for the orbital magnetic moment (see Fig.~\ref{fig:ms_ml}(b)).

\begin{table}[t]
    \centering
    \caption{\label{tab-results}
    Spin and orbital magnetic moments (in $\mu_B$)  on X = B, C, and N in researched systems. The DFT calculations were performed with the full-potential local-orbital (FPLO-18) code \cite{koepernikFullpotentialNonorthogonalLocalorbital1999}. The exchange-correlation functional in the Perdew-Burke-Ernzerhof (PBE) parametrization was used \cite{perdewGeneralizedGradientApproximation1996}. 
    }
    \centering
    \begin{tabular}{cccccccccc}
        \hline \hline
            System &           $m_{(\rm{s},X)}$	&  $m_{(\rm{l},X)}$\\
        \hline
            (Fe$_{0.7}$Co$_{0.3}$)$_{18}$B	&  -0.33 &	0.002 \\
            (Fe$_{0.7}$Co$_{0.3}$)$_{18}$C  &  -0.25 &  0.002 \\
            (Fe$_{0.7}$Co$_{0.3}$)$_{18}$N  &  -0.03 &  0.004 \\
        \hline \hline
    \end{tabular}
\end{table}

At the end, we present the calculation of the total density of states to show what happens after doping (Fe$_{0.7}$Co$_{0.3}$)$_{18}$ thin films with B, C or N, see Fig.~\ref{fig:DOS_1}. 
First, in Fig.~\ref{fig:DOS_1}(a), we show the total DOS for the pure Fe thin film with the comparison to the pure (Fe$_{0.7}$Co$_{0.3}$)$_{18}$ thin film shown in Fig.~\ref{fig:DOS_1}(b).
Compared to pure Fe, (Fe$_{0.7}$Co$_{0.3}$)$_{18}$ has 0.3 more electrons per transition metal atom than pure Fe, causing the Fermi level to shift slightly to the right for the (Fe$_{0.7}$Co$_{0.3}$)$_{18}$ alloy.
Doping the (Fe$_{0.7}$Co$_{0.3}$)$_{18}$ alloy layer with B, C, or N atoms (see Fig.~\ref{fig:DOS_1}(c-e)) results in a small additional contribution of $p$-group electrons, which causes a further shift of the Fermi level to the right. 
This shift is slight because there is one dopant atom for every 18 atoms of $4s$ and $3d$ band atoms in the (Fe$_{0.7}$Co$_{0.3}$)$_{18}$ alloy thin film. 
It can be seen that the dopants form a bond with the alloy, as we do not observe sharp peaks in the total DOS curve.

%
Finally, we would like to comment on the selected limitations of the discussed models resulting from the choice of the size of the system and the usage of virtual crystal approximation.
%
%
The thickness of the 9-monolayers was chosen slightly above the experimentally observed structural transition of ultrathin Fe films from a structure with strong tetragonal deformation ($c/a$ > 1.4) below the critical thickness to a bcc-like structure above it~\cite{cuenya2001observation}. 
At the same time, we did not choose a higher thickness to not increase the number of non-equilibrium atoms in the model and thus the time required for calculations.
Since we did not consider a range of film thicknesses (e.g., from 1 to 20 atomic monolayers), the value of our results for a single unique thickness is limited.
For example, in line with the authors' other research on ultra-thin films~\cite{marciniakL1_0FePt2023, meixner2023magnetic}
the variation in MAE with thickness is not monotonic and can change significantly with a change in thickness by as little as one or two monolayers. 
Hence, the most significant conclusion regarding the MAE of the examined layers is the tendency of MAE to increase with the increase in the atomic number of the dopant.
On the other hand, the calculated magnetic moments on the film's surface, around the dopant, and on it depend not so much on the film thickness.

%
The prepared models also have a specific location of the dopant to the other atoms - it is part of the central layer of the system, creating a very regular distribution of the dopant in it, see Fig.~\ref{fig:structure}.
This assumption was dictated by the need to limit the unit cell size, counted in the number of non-equivalent atoms, to a value enabling MAE calculations.
However, placing dopant atoms in a single monolayer primarily results in overestimating the increase of film thickness due to doping. 
However, as experiments on FeCo thin films with B and C dopants have shown, the magnitude of the tetragonal deformation in our models is relatively consistent with the measurements~\cite{reichel2015lattice, reichel2015soft}.

%
Another approximation used in the presented study is the virtual crystal approximation (VCA), which we applied to model Fe/Co alloying.
We decided to use it even though it is known for overestimating the value of the MAE for bulk materials.
In the case of modeling layered systems containing interstitial defects, which by breaking the symmetry leads to a significant increase in the number of non-equivalent atoms in the system and thus to a significant increase in computational time, the use of the VCA allows both to optimize the system geometry and to reduce the computational time by at least an order of magnitude, relative to alternative methods of (1) averaging over multiple supercells or (2) the coherent potential approximation (CPA), the latter usually happens in combination with the atomic sphere approximation (ASA), leading to unreliable MAE results.

%
Comparing the obtained MAE values with the results from the literature for calculations for bulk systems (see Table~\ref{tab-comparison-MAE-moments}), we notice that the obtained values differ significantly. 
However, these differences may primarily result from the fact that calculations of the layered system are performed, where size effects come into play. 
Even calculations provided using the VCA method in the bulk system resulted in much higher values. 
Comparison with the results of experiments on thin-film systems allows us to confirm the similarity of orders of magnitude.

\section{Summary and Conclusions}

We have conducted a theoretical study of the magnetic anisotropy of ultrathin (9 atomic monolayers thick) FeCo films with B, C, and N dopants located in octahedral interstitial position in the center of the layer. 
The layer structures were subjected to geometry optimization of the interlayer distances and the vicinity region of the dopant sites. This allowed us to get different distances between different layers according to the placement above/below the dopant atom or next to this place and in the different depths of the film thickness. 
We determined the local magnetic moments and excess charge at each position in the films and identified the dopant atoms' effect on the FeCo films' magnetic properties, such as magnetization, total DOS, and magnetic anisotropy.

Contrary to the results for bulk systems, our results indicate that doping FeCo with B, C, and N atoms in the octahedral position in 9 atomic monolayers thick FeCo films can significantly reduce the MAE, even changing its sign to negative in the case of B-doped FeCo thin films. 
The results of this investigation may have important implications for further research on magnetic thin films for spintronic applications.

\section*{Acknowledgements}
We acknowledge the financial support of the National Science Centre Poland under the decision DEC-2018/30/E/ST3/00267.
We thank Pawe\l{} Le\'sniak and Daniel Depcik for compiling the scientific software and administration of the computing cluster 
at the Institute of Molecular Physics, Polish Academy of Sciences.

\bibliography{bib}

\end{sloppypar}
\end{document}